\documentclass[12pt]{article}

\usepackage{amsmath}
\usepackage{amssymb}
\usepackage{latexsym}

\def\beqar {\begin{eqnarray}}
\def\eeqar {\end{eqnarray}}
\def\beq {\begin{equation}}
\def\eeq {\end{equation}}

\def\no2 {{\textstyle{n\over 2}}}

\begin{document}

\rightline{} \vspace{1cm}
\begin{center}

{\Large \bf 4d-flat compactifications with brane vorticities}

\vspace{1.5cm}

S.~Randjbar-Daemi\footnote{The Abdus Salam International Centre
for Theoretical Physics, Trieste, Italy} and V. Rubakov
\footnote{Institute for Nuclear Research of the Russian Academy of
Sciences,
Moscow, Russia}\\

\end{center}
\vskip 0.8cm \centerline{\bf Abstract}
\bigskip
\small{We present  solutions in six-dimensional gravity coupled to
 a sigma model, in the presence of  three-brane sources. The
space transverse to the branes is a compact non-singular manifold.
The example of $O(3)$ sigma
model in the presence of two three-branes
is worked out in detail. We show that the four-dimensional flatness
is obtained with a single condition involving the brane tensions,
which are in general different  and may be both positive, and
another characteristic of the branes, vorticity. We speculate that
the adjustment of the effective four-dimensional cosmological
constant may occur through the exchange of vorticity between the
branes. We then give  exact instanton type solutions for sigma
models targeted on a general K\"ahler
manifold, and elaborate in this framework
on multi-instantons of the $O(3)$ sigma model. The latter have branes,
possibly with vorticities, at the instanton positions, thus
 generalizing our two-brane solution.}

\vspace{3cm}


Theories with extra dimensions offer interesting twists of the
cosmological constant problem~\cite{Rubakov:1983bz}. In
brane-world models, four-dimensional flatness in general does not
require that the tension of the Standard Model brane vanishes. The
effect of this tension on four-geometry may be compensated by the
bulk cosmological constant and/or bulk fields, as well as by
tensions of other branes. It is difficult, however, to invent an
adjustment mechanism for the effective four-dimensional
cosmological constant in those cases, since the parameters
balancing the SM brane tension either are constants of motion or
depend on properties of other branes.

In this paper we present a model of somewhat different sort, with
two compact non-singular extra dimensions
and several branes whose tensions are in general different and
not tuned. Besides the six-dimensional Einstein gravity, our model
involves a scalar field of a non-linear sigma
model. 
Thus, our solutions\footnote{Singular solutions of this type in supergravity models  have been found in 
\cite{Kehagias:2004fb}.
Solutions from the same class as ours 
have been found 
independently in Ref.~\cite{Lee:2004vn}, which appeared after
the arXive version of this paper. The authors of Ref.~\cite{Lee:2004vn}
do not, however, introduce vorticities.}
are
brane-world generalisations of the 
solution found in Ref.~\cite{Gell-Mann:1985if}.
We begin with a simple example of an $O(3)$ sigma model
and a solution which generalizes the flat-space
one-instanton solution of
Ref.~\cite{BP}. In this case the topology of the transverse space
is that of $S^2$, and there are two branes placed at the poles.
A novel property is the vorticity of a brane,
which may be thought of as the Aharonov--Bohm phase of the scalar
field around the brane. In compact transverse space, the
vorticities of the two branes are necessarily equal and opposite.
Four-dimensional flatness requires one relation between the brane
tensions and vorticities. Since the overall vorticity is zero, it
is not inconceivable that the vorticity of each brane may vary in
time due to the vorticity exchange. It is thus tempting to
speculate that the adjustment of the effective four-dimensional
cosmological constant may occur through the exhange of vorticity
between the branes.

In our model both brane tensions may be positive. The compact
transverse space has the shape of distorted two-sphere with
conical defects at the two poles.
 Geometries of this type have been encountered in previous attempts
 to sidestep the cosmological constant
problem~\cite{Carroll:2003db,Navarro:2003vw} (see also
Refs.~\cite{Aghababaie:2003wz, Aghababaie:2003ar,
Gibbons:2003di}). The model which
 has been studied by these authors consisted of the six-dimensional
 Einstein--Maxwell system with a bulk cosmological constant. It was
 shown long time ago that $R^4\times S^2$ with the Maxwell
 field assuming  magnetic monopole configuration is a stable
 solution of this system~\cite{Randjbar-Daemi:1982hi}.
 The authors of
Refs.~\cite{Carroll:2003db,Navarro:2003vw} have shown that the
same
 topology continues  to be a solution in the presence of
delta-function type three-branes too. However, to obtain this, a
certain relationship between the brane tensions and
 the parameters of the action has been
required~\cite{Navarro:2003bf}, which is not the case in our
model.\footnote{For a detailed study of such models see
Ref.~\cite{Burgess:2004dh}.}

We then proceed to general sigma models targeted on arbitrary K\"ahler
manifolds. Remarkably,
we are able to present a class of solutions
to
the system of the scalar field and Einstein
equations in a fairly concrete form.
Making use of this result,
we  give explicit multi-instanton solutions of  $O(3)$ sigma model
coupled to gravity.
Like in the flat case, the instanton positions are moduli of the soultions.
These multi-instantons have branes, possibly with
vorticities, at their centers, thus generalizing
our two-brane solution.

 We start from the action
\[
S= \int \sqrt{-G}\left[M^{-4}R
-\frac{1}{2\lambda^2}\nabla^M\phi^{\alpha}\nabla_M\phi^{\beta}
h_{\alpha\beta}(\phi) + L_{brane}\right]
\]
Here
 $M$ is the six-dimensional
Planck mass and  $\phi^{\alpha}(x)$, in the general case, are real
scalar fields parameterizing a K\"ahler manifold with the metric
 $h_{\alpha\beta}$. For the example of the $O(3)$ model the target
 space is the sphere $S^2$ with the metric $h_{\alpha\beta}$ given by
 \beq
h_{\alpha\beta} = \frac{4}{(1+
\frac{\phi^2}{\alpha^2})^2}\delta_{\alpha\beta} \label{h} \eeq
 where $\phi^2= \phi_1^2 + \phi_2^2$ and  $\alpha$ is the radius of
$S^2$. One can  absorb this parameter into a redefinition
 of $\lambda$; henceforth we  set $\alpha=1$. The six-dimensional
Einstein equations are
\[
R_{MN}- \frac{1}{2}G_{MN}R = M^{-4}T_{MN}
\]
where the energy-momentum tensor is given by
\[
T_{MN}= \frac{2}{\lambda^2}h_{\alpha\beta}(
\nabla_M\phi^{\alpha}\nabla_N\phi^{\beta} -\frac{1}{2}
G_{MN}\nabla^L\phi^{\alpha}\nabla_L\phi^{\beta}) + T_{MN}^{brane}
\]
The brane energy-momentum tensor will be taken to represent a pair
of three-branes parallel to each other. Note that we set the bulk
cosmological constant equal to zero; we shall comment on this
point later on.

 The scalar fields satisfy the highly non-linear equations
\beq
 \nabla^M\nabla_M \phi^\alpha + \Gamma_{\beta\gamma}^\alpha(\phi)
 \nabla^L\phi^{\beta}\nabla_L\phi^{\gamma}=0
\label{+} \eeq
 where $\Gamma$'s are the connection components in the space of
 the $\phi$'s.
    Our ansatz for the solution is
\[
ds^2 = \eta_{\mu\nu}dx^\mu dx^\nu + \psi (y) \delta_{mn} dy^m
dy^n
\]
    where $\mu, \nu =0, 1, 2, 3$ and $y^m$  are the local Gaussian
    coordinates in the two
    dimensional manifold which will support the branes whose world
    volumes are along the $x^\mu$ subspace.
$g^{(2)}_{mn}(y) = \psi (y) \delta_{mn}$ represent
    the components of the metric in the space transverse to the
    branes.

The equations given above are general. Let us consider first  the
example of  $O(3)$ sigma model. The simplest  anzatz for the scalar
fields is
\beq
\phi^\alpha = y^\alpha
\label{phi}
\eeq
 It is straightforward to verify that this ansatz solves the
    scalar field equations (\ref{+})
with no constraints on the parameters
    of the model. More general solutions will be given later on.

To obtain the solution to the Einstein equations, we assume that
the transverse space is symmetric under $O(2)$
rotations, and write its
metric  as follows, \beq
     ds_2^2 = \psi(r)( dr^2 + r^2 d\varphi^2)
\label{*} \eeq where $\psi$ is a function of $r$ only. The
contribution of the scalar fields of the form (\ref{phi}) to
$T_{mn}$ vanishes identically. The only
    non-vanishing components of $T_{MN}$ then become
\[
T_{\mu\nu} = \eta_{\mu\nu} \frac{1}{\psi}
\left[-\frac{8}{\lambda^2}\frac{1}{(1+ r^2)^2}
    -\sum T_i \delta_2 (y-y_i) \right]
\]
The location of the branes with the tension $T_i$ has been denoted
by
    $y_i$.
 The only non-trivial information is contained in the
$\mu\nu$-components of the Einstein equations, which yield \beq
\psi R^{(2)} = \frac{16}{\lambda^2M^4}\frac{1}{( 1+r^2)^2} +
\frac{2}{M^4}\sum T_i \delta^{(2)} (y_-y_i) \label{EE} \eeq
 where $R^{(2)} = - \frac{1}{\psi}\Delta^{(2)} \ln \psi$ is the scalar
 curvature of the two-dimensional transverse space.
Outside the branes, the solution to eq.~(\ref{EE}) is \beq
  \psi(r) = \beta^2
\frac{r^{-\frac{\tau_0}{\pi M^4}}}{(1+ r^2)^{\frac{4}{\lambda^2
M^4}}} \label{soln} \eeq where $\beta$ and $\tau_0$ are yet
undetermined constants. For
\[
      \frac{\tau_0}{2\pi M^4} + \frac{4}{\lambda^2 M^4} > 1
\]
the proper distance from the origin to $r=\infty$, as well as the
volume of the transverse space are finite. The transverse space
thus has the topology of $S^2$. In general, there are conical
defects at $r=0$ and $r=\infty$, i.e., there are branes of
non-vanishing tensions at the two poles.

We still have to specify the range of the coordinate $\varphi$.
Let us choose
\beq
   \varphi \in [0, 2\pi(1- \kappa)]
\label{vort-def}
\eeq
where $\kappa$ is yet another parameter of the model. With this
choice, we introduce the vorticities of the branes placed at $r=0$
and $r= \infty$. Indeed, the scalar field configuration
(\ref{phi}) may be written as
\[
  \phi^1 + i\phi^2 = r \mbox{e}^{i \varphi}
\]
It is not single valued: as the angle $\varphi$ makes full
rotation, i.e., changes from $0$ to $2\pi (1- \kappa)$, the field
obtains the phase factor $\exp (-2\pi i \kappa)$. Yet the physical
quantities like energy-momentum tensor or currents are
single-valued, so this construction makes sense. Physically, it
may be realised, e.g., if the $U(1)$ symmetry $\phi \to
\exp(i\alpha) \cdot \phi$ is gauged (with negligibly small gauge
coupling), and the brane at the origin carries the Aharonov--Bohm
flux proportional to $\kappa$. Clearly, the brane at $r=\infty$
has the vorticity of equal magnitude and opposite sign (e.g., it
carries the opposite Aharonov--Bohm flux).

Near the origin, a coordinate transformation brings the metric
(\ref{*}) to the metric of the 2-dimensional Euclidean plane,
\[
ds^2 = d\rho^2 + \nu^2 \rho^2 d\varphi^2
\]
 where $\nu= 1- \frac{\tau_0}{2\pi M^4}$. Introducing a
new polar angle
 $\varphi'= \nu\varphi$, we recover the standard flat metric
except that  $\varphi'$ ranges from zero
 to $2\pi \nu (1-\kappa)$. We thus obtain a deficit angle
$\delta = 2\pi (1- \nu + \nu\kappa)$. Making use of the standard
relation between the brane tension at the origin, $T_0$, and the
deficit angle, namely, $\delta = \frac{T_0}{M^4}$, we express the
parameter $\tau_0$ of the solution through  the tension and
vorticity
 of the brane placed at the origin,
\beq
  \tau_0 = \frac{T_0 - 2\pi \kappa M^4}{1-\kappa}
\label{tau} \eeq Another parameter of the solution, $\beta$,
remains undetermined; it is thus a modulus.

 As $r$ approaches infinity, the metric becomes
\[
  ds_2^2 = \beta^2 \xi^{-\frac{\tau_\infty}{\pi M^4}}(d\xi^2 +\xi^2
 d\varphi^2)
\]
 where $\xi=\frac{1}{r}$ and $\tau_\infty$ is defined by
\beq
 \tau_\infty  + \tau_0 = 4\pi M^4
\left(1 - \frac{2}{\lambda^2 M^4}\right) \label{sum} \eeq The
relation between $\tau_\infty$ and the brane tension at $r=\infty$
is again given by eq.~(\ref{tau}), with $\tau_\infty$ and
$T_\infty$ substituted for $\tau_0$ and $T_0$, respectively.
Therefore, eq.~(\ref{sum}) is in fact the relation between the
brane tensions and vorticity, for given parameters of the action,
\beq
   T_\infty + T_0 = 4\pi
\left(M^4 - \frac{2 (1-\kappa)}{\lambda^2} \right)
\label{rel}
\eeq
It is this relation that ensures the absense of singularities
and four-dimensional flatness of our solution. As discussed above,
for given $T_0$ and $T_\infty$ it requires the tuning of the
vorticity $\kappa$.

Several remarks are in order.

(i) Clearly, there is a domain of parameter space where the
tensions of both branes can be positive. The vorticity can have
either sign.

(ii) As a cross check, one can calculate the Euler number of the
transverse space, and find that it is equal to
 +2. This reiterates that
our space is  topologically  $S^2$.

(iii) One finds from eqs.~(\ref{EE}) and (\ref{soln})
that the Ricci scalar $R^{(2)}$ vanishes at the brane positions.
We will make use of this property to construct more general solutions
in what follows.

(iv)
 To obtain our solution, we have assumed that the six-dimensional
cosmological constant is zero. It has been shown very 
recently~\cite{Lee:2004vn} that in the model under study, the
non-vanishing bulk cosmological constant leads to
(anti-)~de~Sitter geometry on the branes, i.e., it gives rise to
non-zero effective four-dimensional cosmological constant. 
One way to ensure that the six dmensional cosmlogical constant vanishes
would be to find a six-dimensional
supergravity model, such as the anomaly
 free model of Ref.~\cite{Randjbar-Daemi:wc} where there are
 several scalar fields
 parameterizing a non-linear sigma model manifold, and where the
 six-dimensional
cosmological constant is zero by supersymmetry. Unfortunately
 the target spaces of the sigma models in such supergravities
 are not of the type we have assumed. Nevertheless holomorphic embedding is
 possible, although the most straightforward one yields a singular
 \cite{Kehagias:2004fb}
 transverse space.
 \footnote{After this
 paper appeared in arXive, 
it has been shown in Ref.~\cite{Nair:2004yu} that analogous  
non-singular positive tension branes also exist
 in supergravity models.}

Let us now show that the solution presented above is
the simplest among a wider class of solutions in which the target
space of the sigma model can be any K\"ahler manifold. Such
manifolds are complex with the metric tensor derivable from a
potential $\chi( \phi^a, \phi^{\bar a})$, where $\phi^a$ and
$\phi^{\bar a}$ are local complex coordinates. The only
non-vanishing components of the Hermitean metric are $h_{a\bar b}=
\partial_a\partial_{\bar b}\chi$, where
$\partial_a=\frac{\partial}{\partial \phi^a}$ and
$\partial_{\bar a}$ is the complex conjugate operator. The
 connection components can be calculated by
using the standard formula. The non-vanishing
ones are $\Gamma^a_{bc}= h^{a\bar d}\partial_b h_{\bar
 d c}$ and their complex conjugates.

 It is convenient to introduce
complex coordinates in the transverse $y$-space
 too. Let us denote them by $z$ and $\bar z$. It is
straightforward to see
 that the $\phi$-field equations are satisfied if
\beq
  \partial_{\bar z} \phi^a =0
\label{gen1}
\eeq
  These are  instanton configurations.
 A remarkable fact is that we can solve the
Einstein equations for
  a general K\"ahler manifold. The solution is
\beq
  \psi(z, \bar z) = | g(z)|^2 \exp \left[-\frac{2}{\lambda^2
  M^4}\chi(\phi,\bar \phi) \right]
\label{gen2}
\eeq
  where $g(z)$ is a function of $z$ but not $\bar z$.
It is worth noting that the Euler number
of the transverse space for this solution can be
written in a fairly explcit form as well.
We obtain, using the Einstein equations,
\beq
\frac{1}{4\pi}\int~d^2 x~ \sqrt{g^{(2)}} R^{(2)}
= \frac{2}{\pi \lambda^2M^4}\int~dzd\bar{z}~
  \partial_z \partial_{\bar z}\chi + \frac{1}{2\pi M^4}\sum T_i
\label{Euler}
\eeq
  Here the first integral on the right hand side is the
pull back of the K\"ahler class of the target space of the sigma model.

We leave the analysis of the
properties of these general solutions for future work, and here
we make use of eqs.~(\ref{gen1}) and (\ref{gen2}) to obtain
multi-instanton solutions in  $O(3)$ sigma model.

Let us first consider branes without vorticities. In the case of
  $O(3)$ model the $N$-instanton configuration is given by
\beq
  \phi = \prod_{k=1}^N \frac{z-a_k}{z-b_k}
\label{novort3}
\eeq
  where $a_k$ and $b_k$ are the $2N$ complex moduli of the
  instantons. The solution is normalized in such a way
  that it remains
  finite as $z\rightarrow\infty$.
In other words, we choose the coordinates $z$ in the transverse
space such that there is no brane at $z=\infty$.
For $N=1$
  we obtain our previous solution
  by a simple holomorphic change of the $z$-coordinate.

The potential for the $O(3)$ model is
  $\chi(\phi, \bar \phi) = 2 \ln ( 1+
  |\phi|^2)$. We use the analogy to the one-instanton case,
and determine the function $g(z)$ in eq.~(\ref{gen2}) by the requirement
  that the Ricci scalar $R^{(2)}$ vanishes at $z=a_l$ and $z=b_l$.
This condition fixes $g(z)$ to be
\beq
   g(z)= \beta\prod_1 ^N
\frac{(z-a_l)^{-\frac{\tau_l}{2\pi M^4}}}{(z-b_l)^{-
\frac{\tau_l^\prime}{2\pi M^4}+\frac{2}{N}}}
\label{novort4}
\eeq
   where $\beta$ and $\tau_l$ and $\tau_l'$ are constants. It is
straightforward to
find the metric near the instanton centers,
\beqar
   ds_2^2 &=&  |z-a_l|^{-\frac{\tau_l}{\pi M^4}}~dz d \bar{z}
\; , \;\;\; z \to a_k
\label{*2}
\\
      ds_2^2&=&|z-b_l|^{-\frac{\tilde\tau_l}{\pi M^4}}~dz d\bar{z}
\; , \;\;\; z \to b_k
\label{*3}
\eeqar
where
\beq
\tilde\tau_l= -\tau_l' + \frac{4\pi M^4}{N}\left(1 -
    \frac{2N}{\lambda^2M^4}\right)
\label{novort1}
\eeq
    There are thus deficit angles at each  of the points
    $z=a_l$ and $z=b_l$, which indicate the presence of $2N$
    three-branes sitting at these points.
Their tensions are (recall that we consider branes without
vorticities)
\beqar
          T_l &=& \tau_l \; , \;\;\;\; z=a_l
\label{number}
\\
 \tilde{T}_l &=&\tilde{\tau_l} \; , \;\;\;\; z=b_l
\label{novort2}
\eeqar
In order that the deficit angles and hence the tensions  be
    positive we need to require that both $\tau_l$
    and $\tilde{\tau}_l$
are positive. This will also ensure that the Ricci
    scalar $R^{(2)}$ vanishes
at the position of the branes. Finally,
the transverse space is compact and does not have a conical
singularity at $z=\infty$
provided that $g(z)=1/z^2$ as $z \to \infty$.
This implies a sum rule
\beq
    \sum_1 ^N (\tau_l- \tau_l') =0
\label{sumtau}
\eeq
In view of eqs.~(\ref{novort1}) and (\ref{novort2})
this is in fact a sum rule for the brane tensions, whose
explicit form will be given below.

Let us now introduce the brane vorticities. We
denote the vorticities
around $a_l$ and $b_l$ by $\kappa_l$ and $\kappa'_l$, respectively.
Full rotations around each of the branes should induce the
corresponding phases in $\phi$, so the global solution
for the scalar field, instead of eq.~(\ref{novort3}), now
has the form
\beq
 \phi = \prod_{k=1}^N \frac{(z-a_k)^{1-\kappa_k}}{(z-b_k)^{1+\kappa^\prime_k}}
\label{vort3}
\eeq
To make contact with the definition (\ref{vort-def}), we note that
near $z=a_k$, the coordinate transformation $y=(z-a_k)^{1-\kappa_k}$
brings the field (\ref{vort3}) into the form (\ref{phi}), while the
phase of $y$ ranges from zero to $2\pi (1-\kappa_k)$, in accord with
eq.~(\ref{vort-def}). Similar remark applies to branes at sites $b_k$.

The field $\phi$ should tend to a constant as $z \to \infty$ (no brane
at $z=\infty$), which
implies  that the sum of vorticities vanishes,
\beq
    \sum_{l=1}^N (\kappa_l + \kappa^\prime_l) = 0
\label{sumvort}
\eeq
Now, the metric is still given by eqs.~(\ref{gen2}) and
(\ref{novort4}), and its behaviour near the instanton centers
is again determined by eqs.~(\ref{*2}) and (\ref{*3}).
However, instead of eq.~(\ref{novort1}) we now have
\[
\tilde\tau_l= -\tau_l' + \frac{4\pi M^4}{N}\left[1 -
    \frac{2N}{\lambda^2M^4}(1+\kappa_l^\prime)\right]
\]
The relations (\ref{number}) and
(\ref{novort2}) are still valid, so
the sum rule (\ref{sumtau}) gives one relation between
the tension and vorticities,
\[
  \sum_l (T_l + \tilde{T}_l) =
4\pi \left[ M^4 - \frac{2}{\lambda^2} \left(N - \sum_l \kappa_l\right)
\right]
\]
where we used the sum rule (\ref{sumvort}). This generalizes the
relation (\ref{rel}) to the multi-brane case. Making use of
eq.~(\ref{Euler}) one checks that this solution has
the Euler number +2.

To end up this note,
let us point out a particular case when  $a_1=a_2=...=a_N \equiv a$ and
    $b_1=b_2=...=b_N \equiv b$. In this case we have
    essentially a single brane at $z=a$ and another one at $z=b$. By
    the coordinate transformation of the form
    $\xi=\frac{z-a}{z-b}$,
the formulae given in the last paragraph reduce to the ones
    very similar to the one instanton example, except that we need
    to replace $r$ in the $\psi$-function by $r^N$, where
    $r=|\xi|$. The relationship (\ref{rel}) between the tensions
    is then  replaced by
    \[
   T_\infty + T_0 = 4\pi
\left(M^4 - \frac{2N (1-\kappa)}{\lambda^2} \right) \label{rel'}
\]
 This shows that even for $\kappa=0$, at large $\lambda$
we obtain a dense,
albeit discrete, set  in the space
 of tensions, which is parameterized by the instanton number $N$.

\section*{Acknowledgements}

V.R. thanks the Abdus Salam International Centre for Theoretical
Physics, where part of this work has been done, for hospitality.
This work was supported in part by RFBR grant 02-02-17398.

\end{document}